\documentstyle[twoside,fleqn,espcrc2,epsfig]{article}

\newcommand{\beq}{\begin{eqnarray}}
\newcommand{\eeq}{\end{eqnarray}}

\newcommand{\la}{\langle}
\newcommand{\ra}{\rangle}
\newcommand{\qc}{\la \bar{q}q \ra}
\newcommand{\uc}{\la \bar{u}u \ra}
\newcommand{\dc}{\la \bar{d}d \ra}

\newcommand{\dmdmu}{{\partial m \over \partial \mu}}

\newcommand{\dmkdmu}{{\partial m_K \over \partial \mu}}
\newcommand{\dmkdmus}{{\partial m_K \over \partial \mu_S}}

\newcommand{\dmpidmu}{{\partial m_\pi \over \partial \mu}}

\newcommand{\ddqcdmu}{{\partial^2 \qc \over \partial \mu^2}}
\newcommand{\ddmdmu}{{\partial^2 m \over \partial \mu^2}}

\begin{document}

\title{Chemical potential response of pseudoscalar meson masses
in the Nambu--Jona-Lasinio model \thanks{In memory of Prof. Osamu
Miyamura, a very kind and warm-hearted person as well as a great
physicist.}}
\author{O.~Miyamura\address[1]{Dept. of Physics, Hiroshima University,
                Higashi-Hiroshima 739-8526, Japan} and
        S.~Choe\addressmark[1]}
\begin{abstract}
Using the Nambu--Jona-Lasinio (NJL) model we study chemical
potential response of the pion and kaon masses as a function of
temperature and chemical potential, i.e., $\dmdmu$($T$, $\mu$).
First, we obtain the responses assuming the vector--axial vector
coupling is zero ($g_V$=0). Then, we include a non-zero $g_V$ and
study the effects of $g_V$ on the responses.
 We find that the behavior of $\dmdmu$ for
 the pion is quite different
 from that for the kaon. It means that $\dmdmu$ is much dependent
 on the mass difference between the two quarks, i.e.,
 the $u$ and $s$ quarks (or even between the $u$ and $d$ quarks).
 Our results may give a clue to future studies
of $\dmdmu$ on the lattice.
\end{abstract}
\maketitle

\section{Introduction}

While the structure of QCD at finite temperature has been
investigated in detail using lattice QCD, little is known about
matter at finite baryon density due to the sign problem
\cite{sh01}. In order to avoid this difficulty one can consider
2--color QCD \cite{sk01,mnn01}, imaginary chemical potential
simulations \cite{hlp01}, the response of a hadron mass to changes
in the chemical potential, $\dmdmu$ and $\ddmdmu$, at zero
chemical potential ($\mu$=0) \cite{taro1}, and so on.

Since the direct application of QCD at finite temperature and
chemical potential is not available in the present lattice QCD
simulations, effective models of QCD are commonly used. One of the
most popular models is the Nambu--Jona-Lasinio (NJL) model
\cite{njl}. This model has been widely used for describing hadron
properties in hot and/or dense matter \cite{hk94}.

In this work we present NJL model calculations of $\dmdmu$ for the
pion and kaon. The primary goal of our study is to get the same
quantity which is simulated on the lattice, i.e., $\dmdmu$($T$,
$\mu$=0). Of course, the direct comparison between the lattice
data and the present NJL model calculations is rather difficult
because $\dmdmu$ on the lattice means the variation of a screening
mass, while it corresponds to the variation of a dynamical mass in
the present work. Nevertheless, we can compare both results to
each other qualitatively.

In contrast to the lattice simulations, we also get $\dmdmu$ at
finite chemical potential within this model. Then, this may give
information on the role of the light quark (the $u$ and $d$
quarks) chemical potential and/or the strange quark chemical
potential in hot and/or dense matter. In addition, we consider two
cases for the pion. One is the pion with the same $u$, $d$ quark
mass. The other is the pion with non-degenerate $u$ and $d$ quark
masses. Since the present lattice QCD can not simulate the
characteristics coming from slightly different $u$ and $d$ quark
masses \cite{tf01}, the response of the pion mass with different
$u$ and $d$ quark masses can be useful for future studies of
$\dmdmu$ on the lattice.

Due to the limitation of space we present only the results with a
non-zero vector--axial vector coupling $g_V$ in this paper. Please
refer to \cite{mc1} for basic formulae to get $\dmdmu$ for the
kaon and pion in the NJL model and the results without $g_V$.
Detailed derivations including small corrections in the figures in
\cite{mc1} will be given elsewhere \cite{mc2}.

\section{The effects of a non-zero $g_V$}
\label{sec2}

\begin{figure}[t!]
\center{\epsfig{file=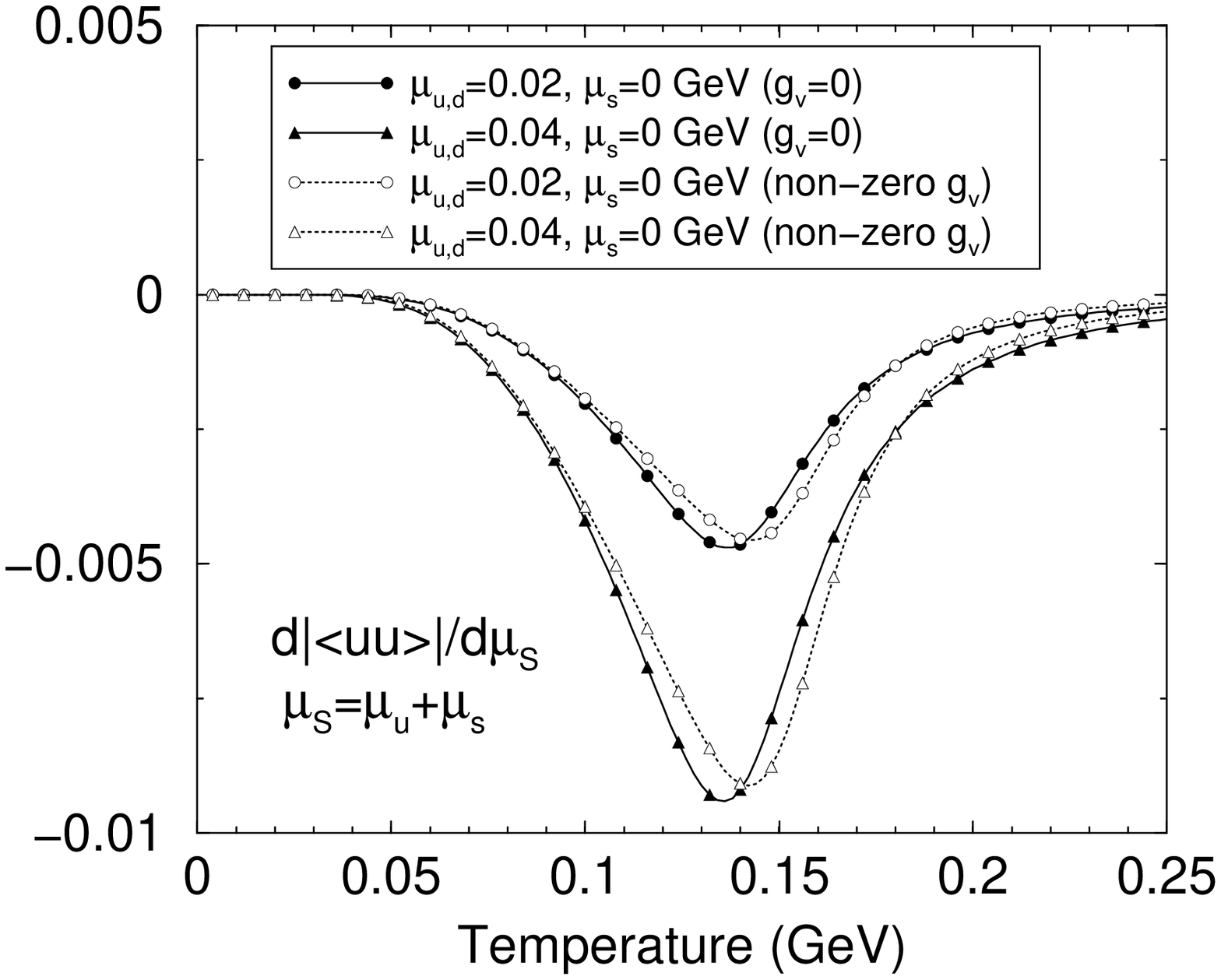,height=6cm,width=7cm}}
 \vspace{0.5cm}
 \center{\epsfig{file=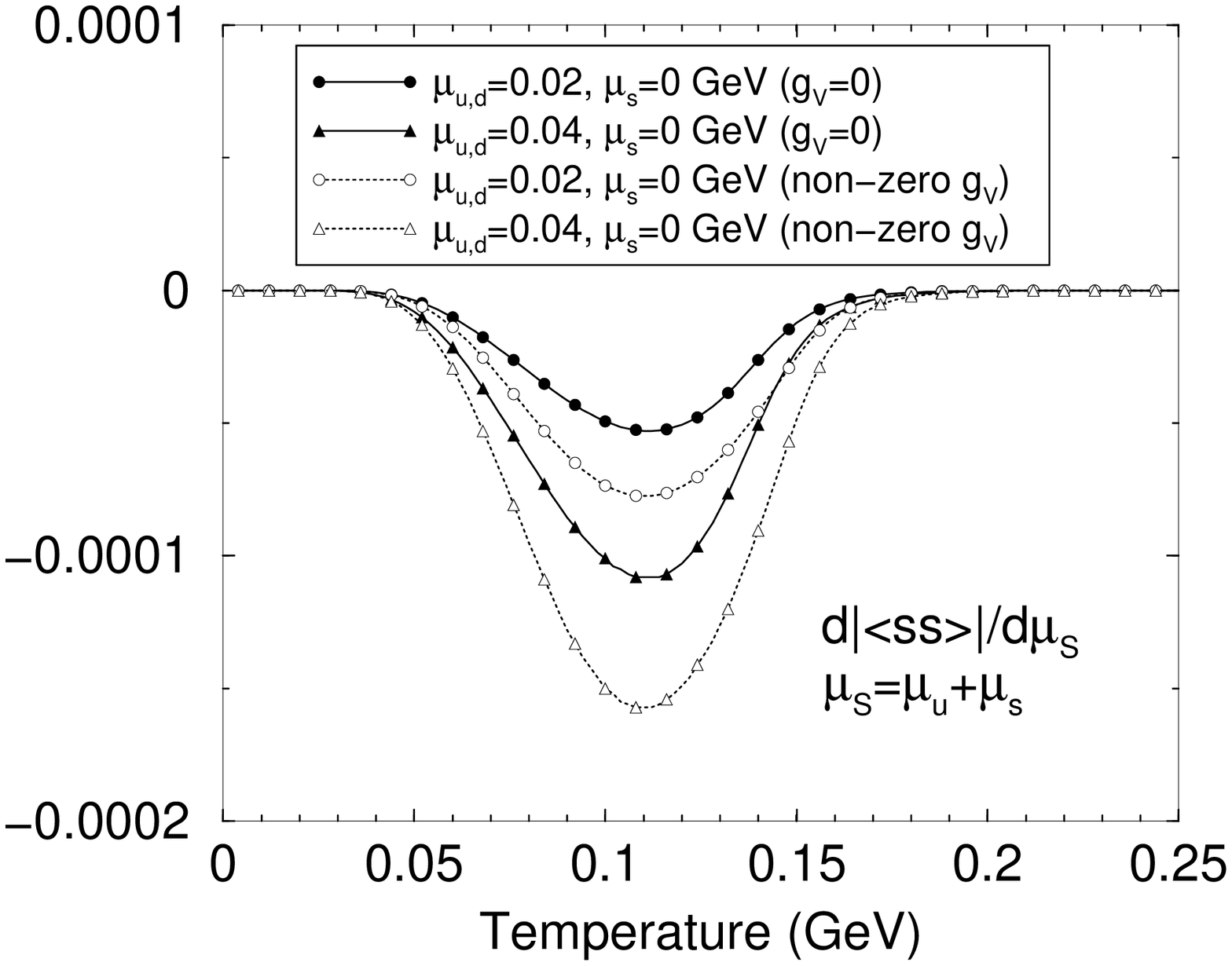,height=6cm,width=7cm}}
 \caption{The responses of quark condensates without $g_V$
 (filled symbols) and with a non-zero $g_V$ (open symbols):
  the $u$ quark condensate (upper) and the $s$ quark
condensate (lower).}
 \label{dqc-gv}
\end{figure}
\begin{figure}[t!]
\center{\epsfig{file=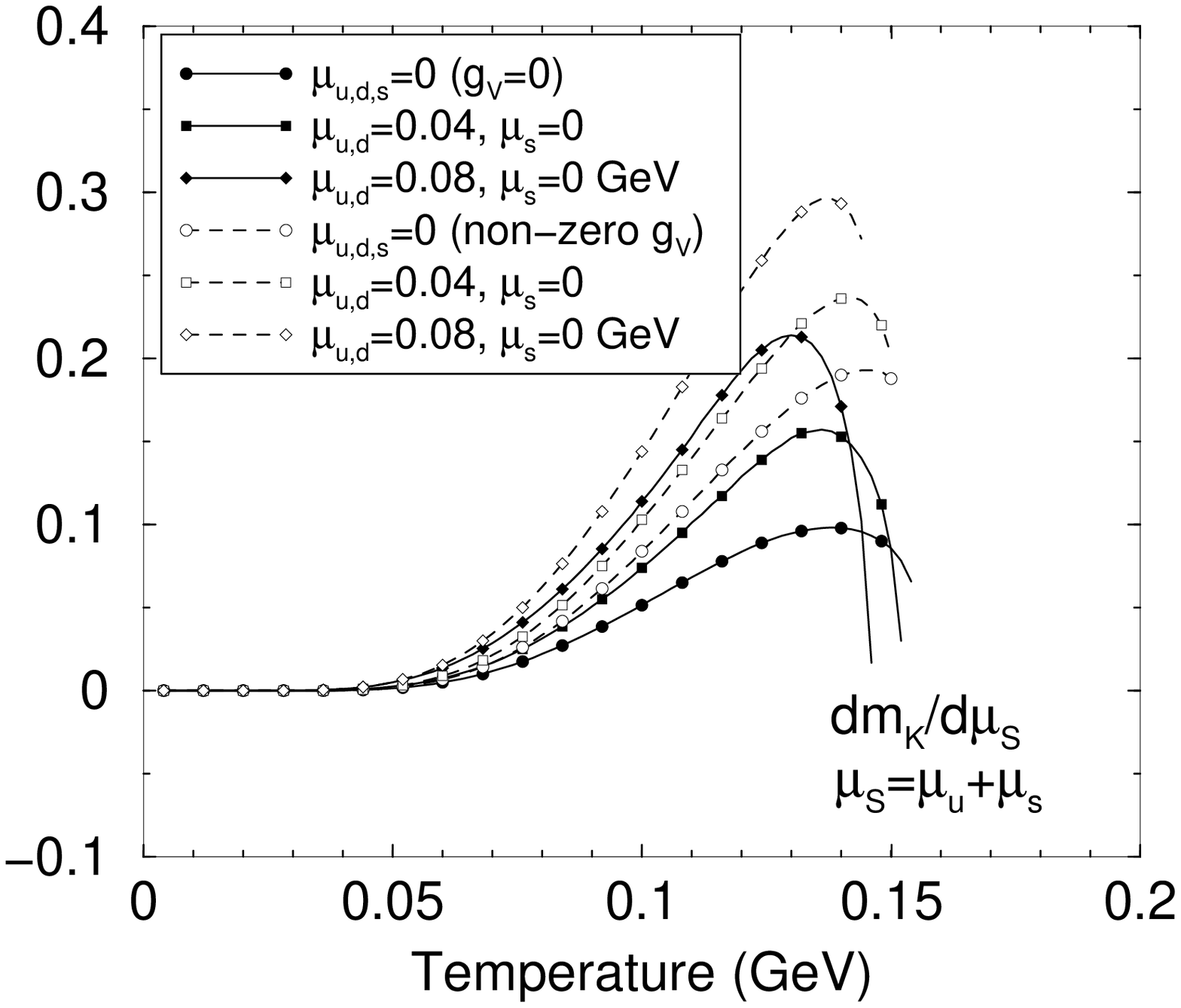,height=6cm,width=7cm}}
 \vspace{0.5cm}
 \center{\epsfig{file=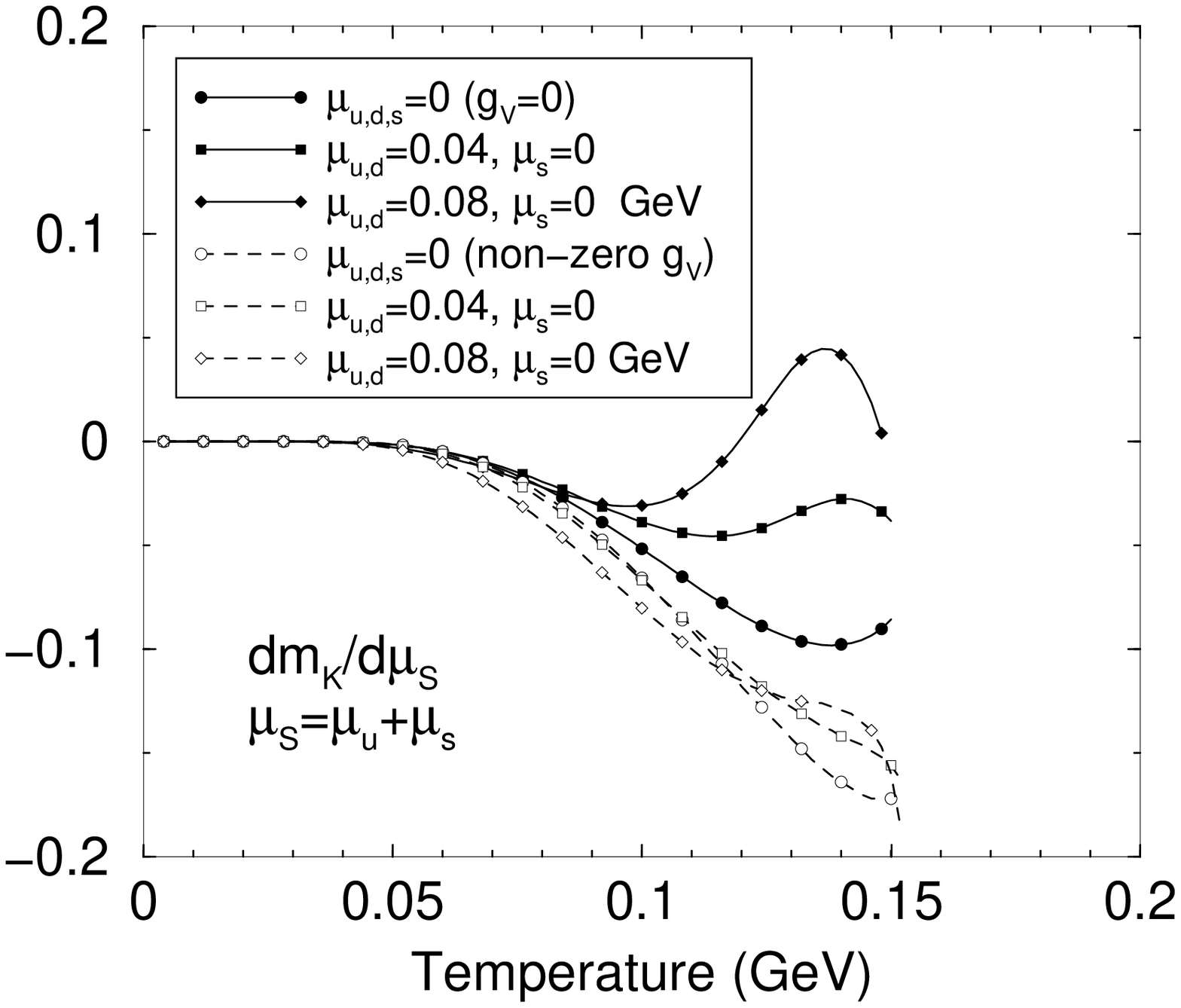,height=6cm,width=7cm}}
 \caption{$\dmkdmus$ without $g_V$ (filled symbols) and
 with a non-zero $g_V$ (open symbols): $K^+$ (upper) and $K^-$ (lower).}
 \label{isos-k-gv}
\end{figure}
One of NJL model calculations showed that the $K^-$ mass at finite
density with $g_V\neq$0 is quite different from that with $g_V$=0
\cite{rsp99}. In this section, we show the chemical potential
response of the kaon mass by including a non-zero $g_V$. In the
case of the pion, the effect of the non-zero vector--axial vector
coupling is negligible \cite{mc2}.

We follow the formalism and the same parameters in \cite{rsp99}.
First, in Fig. \ref{dqc-gv} let us show the responses of quark
condensates with (and also without) $g_V$ at $\mu_u=\mu_d$=0.02,
0.04 GeV, respectively. The effect of a non-zero $g_V$ on the $s$
quark condensate is much larger than that on the $u$ quark
condensate. This behavior can be understood from a large
difference in the $s$ quark mass, i.e., $m_s$=135.7 MeV in
\cite{mc1} and $m_s$=88.0 MeV in the present work. Since the
response of the $s$ quark condensate becomes much larger than
before, one can expect that $\dmkdmu$ for the kaon will be quite
different from the previous values.

In Fig. \ref{isos-k-gv}, we show $\dmkdmus$ for $K^+$ and $K^-$.
For comparison we also show the previous results without $g_V$.
The comparison of $\dmkdmus$ at each point is rather meaningless
because all the parameters including the quark masses are changed.
We would like to show only a qualitative behavior of $\dmkdmus$ in
this figure. The mass of $K^+$ is slightly affected by the
vector-axial vector interaction. On the other hand, the behavior
of $\dmkdmus$ for $K^-$ is much different from the previous one,
where $g_V$=0. This result confirms the previous NJL model
calculations that the vector--axial vector interaction reduces the
effects of the Fermi sea and the $K^-$ mass is a smoothly
decreasing function of density \cite{lsw92,rsp99}.

\section{Concluding Remarks}
\label{sec3}

Using the NJL model we have calculated the chemical potential
responses of the kaon and pion masses, $\dmkdmu$ and $\dmpidmu$,
at zero and finite chemical potential, and found that their
behaviors are quite different from each other. Our results show
that $\dmdmu$ is much dependent on the mass difference of two
quarks, i.e., the mass difference between the $u$ and $s$ quarks,
or even between the $u$ and $d$ quarks. Even at very small
chemical potentials $\dmkdmu$ is much different from $\dmpidmu$.

Let us discuss some uncertainties in our calculations. First, the
mass in our formalism means the dynamical mass, while the
screening mass is simulated on the lattice. For the direct
comparison we may need the formalism to calculate the screening
mass in the NJL model as given in \cite{ff9496}. This may help us
to properly interpret the lattice data. Second, while the lattice
simulations of $\dmdmu$ show a large difference between in the
confinement phase and in the deconfinement phase \cite{taro1}, we
can not predict $\dmdmu$ above the Mott temperature in the present
work. This is because we have excluded the imaginary part of the
1--loop polarization function in the dispersion relation and
considered only the region below the Mott temperature
\cite{mc1,mc2}. Thus, the imaginary part of the polarization
function should be included for a detailed analysis above the Mott
temperature.

As a final remark, we confirm that {$\partial^2 |\qc| \over
\partial \mu^2$} is negative within the present NJL model approach,
and find a clear quark mass dependence between {$\partial^2 |\uc|
\over \partial \mu^2$} and {$\partial^2 |\dc| \over \partial
\mu^2$} both below and above the Mott temperature \cite{mc2}. The
second order response of the quark condensate $\ddqcdmu$ at
$\mu$=0 can also be simulated on the lattice \cite{taro2}, and its
behavior is similar to the result from the present NJL model
calculation.

\section*{Acknowledgements}

 We thank A. Nakamura for valuable comments.
The work of S.C. was supported by the Japan Society for the Promotion
of Science (JSPS).


\end{document}